\documentclass[a4paper]{jpconf}
\usepackage[dvips]{graphicx}
\usepackage{txfonts}
\usepackage{amsfonts,amssymb}
\usepackage{color}
\usepackage{times}
\definecolor{orange}{rgb}{1,0.5,0}
\newcommand{\be}{\begin{equation}}
\newcommand{\ee}{\end{equation}}
\newcommand{\bea}{\begin{eqnarray}}
\newcommand{\eea}{\end{eqnarray}}
\begin{document}

\title{Stellar Scattering and the Formation of Hot-Jupiters in Binary Systems}
\author{J G Mart\'{\i} and C Beaug\'e}
\address{Instituto de Astronom\'ia Te\'orica y Experimental, Observatorio Astron\'omico, Universidad Nacional de C\'ordoba, Laprida 854, (X5000BGR) C\'ordoba, Argentina}
\ead{javi@oac.uncor.edu}

\begin{abstract}
Hot Jupiters (HJs) are usually defined as giant Jovian-size planets with orbital periods $P \le 10$ days. Although they lie close to the star, several have finite eccentricities and significant misalignment angle with respect to the stellar equator, leading to $\sim 20\%$ of HJs in retrograde orbits. More than half, however, seem consistent with near-circular and planar orbits.

In recent years two mechanisms have been proposed to explain the excited and misaligned sub-population of HJs: Lidov-Kozai migration and planet-planet scattering. Although both are based on completely different dynamical phenomena, at first hand they appear to be equally effective in generating hot planets. Nevertheless, there has been no detailed analysis comparing the predictions of both mechanisms, especially with respect to the final distribution of orbital characteristics.

In this paper we present a series of numerical simulations of Lidov-Kozai trapping of single planets in compact binary systems that suffered a close fly-by of a background star. Both the planet and the binary component are initially placed in coplanar orbits, although the inclination of the impactor is assumed random. After the passage of the third star, we follow the orbital and spin evolution of the planet using analytical models based on the octupole expansion of the secular Hamiltonian. We also include tidal effects, stellar oblateness and post-Newtonian perturbations. 

The present work aims at the comparison of the two mechanisms (Lidov-Kozai and planet-planet scattering) as an explanation for the excited and inclined HJs in binary systems. We compare the results obtained through this paper with results in Beaug\'e \& Nesvorn\'y 2012, where the authors analyze how the planet-planet scattering mechanisms works in order to form this hot Jovian-size planets.

We find that several of the orbital characteristics of the simulated HJs are caused by tidal trapping from quasi-parabolic orbits, independent of the driving mechanism (planet-planet scattering or Lidov-Kozai migration). These include both the 3-day pile-up and the distribution in the eccentricity vs semimajor axis plane. However, the distribution of the inclinations shows significant differences. While Lidov-Kozai trapping favors a more random distribution (or even a preference for near polar orbits), planet-planet scattering shows a large portion of bodies nearly aligned with the equator of the central star.  This is more consistent with the distribution of known hot planets, perhaps indicating that scattering may be a more efficient mechanism for producing these bodies.

\end{abstract}

\section{Introduction}\label{sec1}

More than one hundred of Hot Jupiters (HJ) are presently known around main sequence stars. Although there is no precise definition, for our purposes we will include in this group those  planets with observed masses $m > 0.8 m_{\rm Jup}$ and orbital periods $P \le 10$ days. The lower limit for the mass may appear arbitrary, and is dictated more by dynamical considerations than by physical properties of planetary bodies. The upper limit on orbital periods, however, is more easily justified, but also dynamical in nature. For instance, around solar-type stars, this period corresponds to a limit which giant planets in circular orbits suffer significant tidal effects.

The origin of these planets is still a matter of debate. It appears very unlikely that they formed in-situ (e.g. Lin et al. 1996), so their present location must have been achieved after a significant orbital decay from outside the ice line. Although several evolutionary mechanisms were proposed, including disk-planet interactions (e.g. Lin et al. 1996, Ben\'itez-Llambay et al. 2011) and planet-planet scattering (e.g. Rasio \& Ford 1996, Juric \& Tremaine 2008), a smooth planetary migration due to disk-planet interactions appeared as the best candidate. 

Until fairly recently, all detected HJs were consistent with (the assumption of) circular orbits and, more importantly, with values for the missalignment angle consistent with aligned systems (see Winn et al. 2010). These orbital characteristics are expected from disk-induced migration, which led further credibility to this scenario. In the past few years, however, the picture changed. A larger population of HJs was analyzed for the so-called Rossiter-McLaughlin effect, leading to a large portion of bodies displaying significant vaules of the misalignment angle (currently $\sim 40 \%$), including about $\sim 15 \%$ of planets in retrograde orbits with respect to the stellar spin. So, instead of having a rather simple and ``cold'' population of HJs, we are now faced with a more complex dynamics. 

Since significant misalignment angles are not consistent with smooth planetary migration, their origin must lie elsewhere. At this point we are faced with two questions: (i) what other driving mechanism could explain highly inclined and even retrograde planets, and (ii) are all HJs consistent with this new scenario, or must we assume two separate populations of HJs?


Tidal effects tend to align orbits, so any observed misalignment of the orbits must have been caused by the migration mechanism itself. This speaks of a high excitation mechanism which must have affected both the inclination and eccentricity, although the latter may have been later damped by tides. Two scenarios have been proposed for such a mechanism: Lidov-Kozai trapping with a binary companion (e.g. Naoz et al. 2011, 2012), and planet-planet scattering within an initially cold but dynamically unstable planetary system (Nagasawa et al. 2008, Nagasawa \& Ida 2011, Beaug\'e \& Nesvorn\'y 2012). 

Although both scenarios are completely different, the end result is the same. An initially circular orbit of a giant planet beyond the HJ region is excited to high eccentricities (usually close to parabolic orbits) in such a away that the pericentric distance is so close to the star that tidal effects are not only significant but dominant over the gravitational perturbation that generated the excitation. If this process also affected the inclination, then the subsequent orbital evolution of the planet would damp the eccentricity and semimajor axis, leaving as a final product a Hot planet with highly inclined but near-circular orbit. 

In Beaug\'e \& Nesvorn\'y (2012) we showed that planet-planet scattering may explain many of the observed orbital characteristics of HJs, including the eccentricity-semimajor axis distribution, the so-called 3-day pile-up, and the distribution of misalignment angles. Obviously the result depends on the tidal model and the adopted values for the tidal parameters, but this may actually serve as observational constraints on these little-known parameters. 

Although Lidov-Kozai trapping also proved an efficient mechanism, and also explains the existence of highly misalignment HJs, there has been a significantly less comparison with the observed HJ population. For example, it is not clear that this scenario explains the 3-day pile-up, or the eccentricity distribution. In short, which of the observed orbital characteristics are due to the excitation mechanism and which to the subsequent tidal evolution? 

In this work we wish to address precisely these issues. The main idea is to explore the Lidov-Kozai model using similar tools and dynamical models as developed in Beaug\'e \& Nesvorn\'y (2012), changing planet-planet scattering by Kozai resonance. Our aim is then to present a consistent comparison between the predictions of both scenarios, and try to deduce which observed characteristics of the planets are robust and which are model-dependent.

\section{Numerical Simulations}\label{sec2}

Let $m_p$ be a planet orbiting a star $m_A$ and perturbed by its secondary companion $m_B$. We will denote by $a$ the semimajor axis, $e$ the eccentricity, $I$ the inclination, $M$ the mean anomaly, $\omega$ the argument of pericenter and $\Omega$ the longitude of the ascending node.
As with the masses, the orbital elements of the planet will be identified by a subscript $p$, while those of the binary companion by $B$. We will assume Jacobi coordinates with $m_A$ as the central mass, and the reference plane perpendicular to the total orbital angular momentum. 

The equator of $m_A$ will coincide with the orbital reference plane, so the inclination $I_p$ will be equal to the misalignment angle. Finally, we will denote by $I_{\rm mut}$ the mutual inclination between the orbits of $m_p$ and $m_B$, and fundamental to the orbital evolution of the planet due to the gravitational perturbations of the secondary star.

\begin{figure}[t!]
\centerline{\includegraphics*[width=12.0cm]{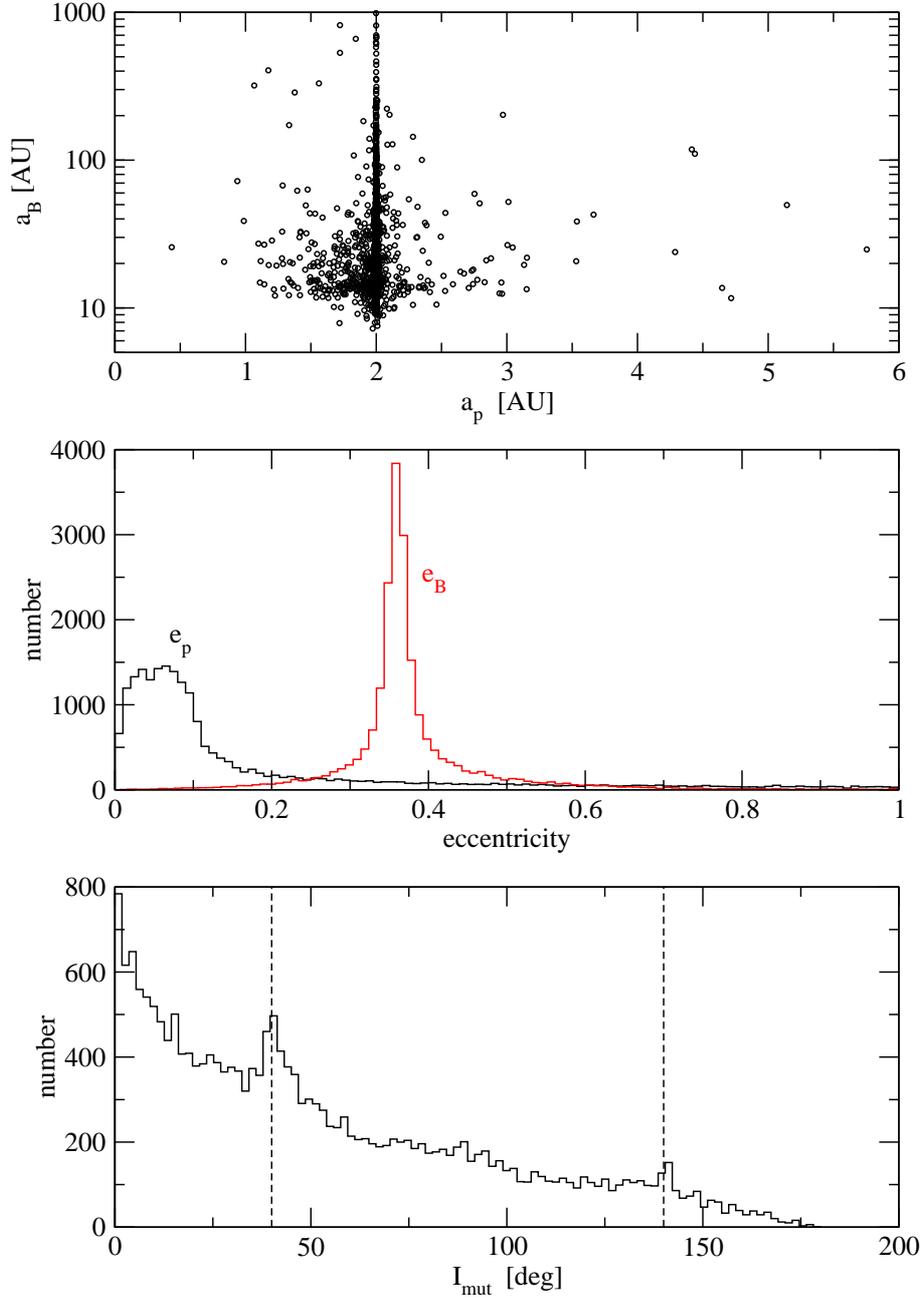}}
\caption{Post scattering distribution of the orbits of both the planet $m_p$ and binary companion $m_B$. The vertical dashed lines in the bottom frame correspond roughly to the limits of the Lidov-Kozai resonance domain.}
\label{fig1}
\end{figure}

\subsection{Initial Conditions}

The simulations presented by Naoz et al. (2011, 2012) started with a planet in circular orbit and aligned with the central star, while the mass and orbit of the secondary was chosen randomly from a predefined distribution. Here we adopted a different route. 


In Mart\'i \& Beaug\'e (2012) we presented a formation scenario for planets in tight binary systems based on the concept of stellar scattering. Initially we assumed a planet with $m_p = 0.002 \textrm{M}_{\odot}$ in a circular orbit with $a_p=2$ AU around a central star of mass $m_A = 1.59 \textrm{M}_{\odot}$, plus a secondary star of mass $m_B = 0.4 \textrm{M}_{\odot}$ with $a_B = 18.5$ AU and $e_B=0.36$. The system was then perturbed by a hyperbolic fly-by with a third star $m_C$ with different initial conditions. This generated a sequence of new binary systems in which both the planet and the secondary star acquired new orbital elements.

Although in that work we focused our interest in reproducing the observed $\gamma$-Cephei system, the result of our simulations generated a series of synthetic binary systems with different orbital characteristics. We will use those end-products of our stellar scattering experiments as initial conditions for our tidal trapping experiments.

We chose a total of 21226 scattering experiments using an impactor mass of $m_C=0.4 m_\odot$, 
a value similar to the mean of the stellar initial mass function for stars (see Kroupa 2001). However, we also checked other values and found no appreciable change in the results. The minimum distance between the impactor $m_C$ and the binary was taken randomly between $0.1$ and $4.0$ times the apocentric distance between both primary stars. Figure \ref{fig1} shows the post-scattering distribution of both the planet and the binary companion which will be taken as initial conditions for our tidal simulations. While most of the planets and binaries retained semimajor axes near their original values, there is a considerable spread in both orbits. The same is also noted in the eccentricities, where $e_B$ has a very narrow distribution around $\sim 0.36$ and practically all values of $e_p$ are restricted to $< 0.2$. 

Finally, while most values of $I_{p}$ remained close to zero, the mutual inclinations $I_{\rm mut}$ show a wider range. Approximately $53 \%$ have values $I_{\rm mut} < 40^\circ$ and thus below the approximate limit necessary for the appearance of the Lidov-Kozai resonance (e.g. Kozai 1962, Harrington 1968, Innanen et al. 1997, Libert \& Henrard 2007, Haghighipour 2010). However, the exact limiting inclination for the resonance is a function of the semimajor axis ratio (Libert \& Henrard 2007), so the value adopted here is illustrative but not rigorous. Close to $40 \%$ of the initial conditions have $40^\circ < I_{\rm mut} < 140^\circ$ and may be affected by the secular resonance. Lastly, near $7 \%$ have highly retrograde orbits, with $I_{\rm mut} > 140^\circ$. Notice that there seems to be two peaks in the distribution of the mutual inclinations (identified by vertical dashed lines), each associated with the limiting values for the appearance of the Lidov-Kozai libration domain.

\begin{figure}[t!]
\centerline{\includegraphics*[width=12.0cm]{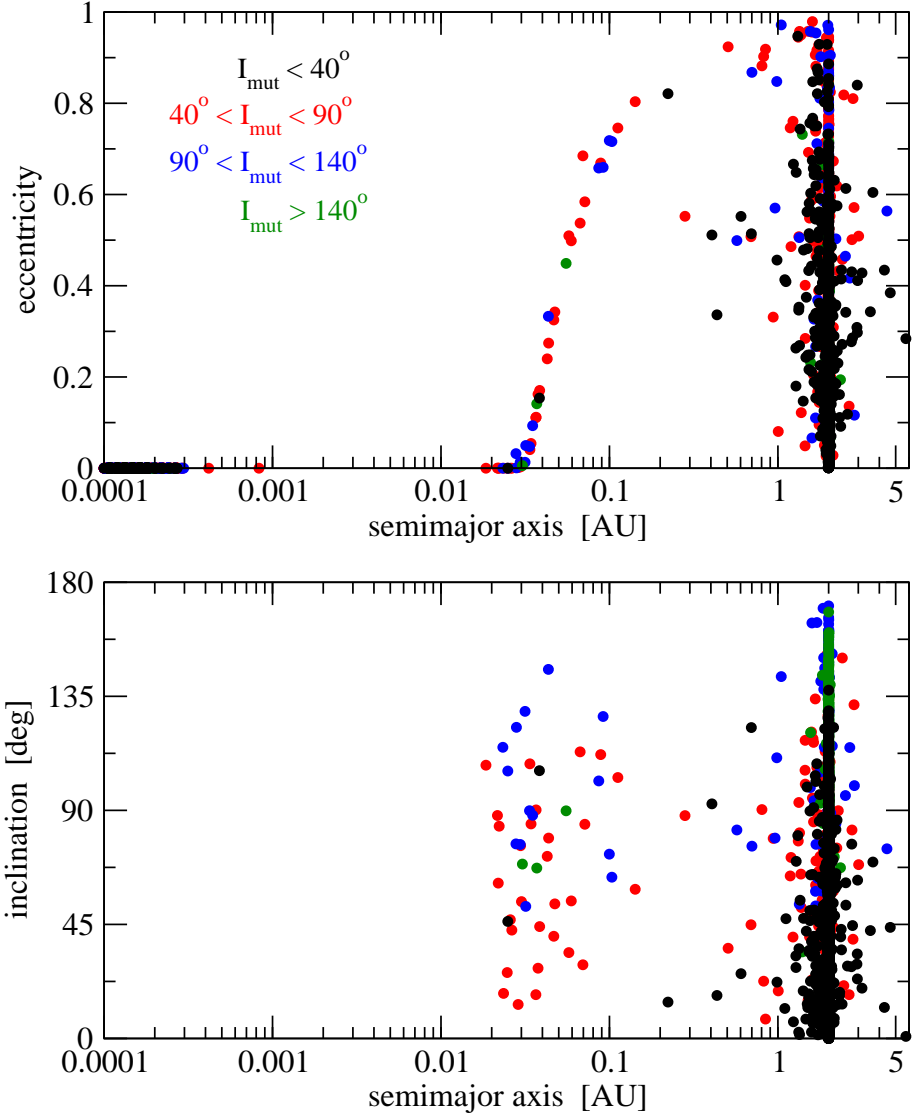}}
\caption{Final distribution of 21226 initial conditions evolved for $1$ Gyr including gravitational perturbations, tidal effects, GR and stellar oblateness. Color code indicates the original mutual inclination between the planet and the secondary star. The Roche radius for the central star is located at $ \sim 0.009$ AU. Most of the tidally trapped bodies do not survive and are engulfed by $m_A$.}
\label{fig2}
\end{figure}

\subsection{Equations of Motion}

This set of 21226 initial conditions were numerically integrated for the total time span of $10^9$ years. The gravitational effects were modeled with an octupole secular Hamiltonian following the description by Laskar \& Bou\'e (2010). We also included tidal effects according to the modified model presented in Beaug\'e \& Nesvorn\'y (2012), and adopted the same values for the tidal parameters: $Q'_p = 5 \times 10^6$ and $Q'_* = 10^7$. The tidal equations contained both the dissipative terms and the precessional effects due to the deformation of the extended body. Finally, for completeness, we also included general relativity effects and stellar oblateness. The expression for both may be found in Beaug\'e \& Nesvorn\'y (2012). 

Together with the orbital elements, we also integrated the spin equations, using the same descriptions as in Correia et al. (2011). The initial rotational periods where chosen equal to $28$ days (for the star) and $10$ days (for the planet). This last value is not really sensitive, since a pseudo-synchronization of the planet was achieved early in the simulation. The initial obliquity of the star was taken equal to zero, and thus the equator coincided with the pre-scattering orbits of both the planet and binary companion. The obliquity of the planet, however, was chosen equal to its inclination with respect to the central star. 

All simulations were stopped if the planet reached the Roche radius of the central star, which corresponds approximately to $r_{\rm Roche} \simeq 0.009$ AU for the cases studied here.

\subsection{Results}

Figure \ref{fig2} shows the final distribution of planetary orbits at the end of the simulations. The top frame shows the distribution of eccentricities as a function of the semimajor axis, where the color code indicates the initial value of $I_{\rm mut}$. Most of the runs lead to very close encounters with the central star and to pericentric distances closer than the Roche radius. Only a few of these still survived in the system after $1$ Gyr, and are shown in the plot. The bottom frame shows the final inclination $I$ with respect to the stellar equator, where we only show those runs with final $a_p > r_{\rm Roche}$. Additional information is shown in Table \ref{tab1}, which indicates the total number of initial conditions in each interval of $I_{\rm mut}$, as well as their outcomes.

\begin{table}
\begin{center}
\begin{tabular}{lrrrr}
\hline
 $I_{\rm mut}$            & Number & Lost & HJs & Others    \\ 
\hline
  $I_{\rm mut} < 40^\circ$             & $7705$  &   $3.2 \%$  &  $0.05 \%$  & $96.7 \%$  \\ 
 \hspace*{0.02cm} $40^\circ < I_{\rm mut} < 90^\circ$  & $3871$  &  $85.1 \%$  &   $0.74 \%$  & $14.1 \%$  \\ 
  $90^\circ < I_{\rm mut} < 140^\circ$ & $3002$  &  $92.5 \%$  &   $0.66 \%$  &  $6.8 \%$  \\ 
 $140^\circ < I_{\rm mut}$             & $1276$  &  $11.8 \%$  &   $0.23 \%$  & $87.9 \%$  \\ 
\hline
\end{tabular}
\end{center}
\caption{Outcomes of the simulations, as function of the initial mutual inclination.}
\label{tab1}
\end{table}

Practically all runs with initial inclinations $I_{\rm mut} < 40^\circ$ (black circles) were not tidally trapped and remained with semimajor axes close to their original values. About $\sim 3 \%$, however, were lost due to engulfment by the central star and only $\sim 0.05 \%$ gave origin to HJs. For $40^\circ < I_{\rm mut} < 90^\circ$, most of the initial conditions were susceptible to excitation by the Lidov-Kozai resonance. The end results show precisely this effect, where only $\simeq 14 \%$ of the bodies remained close to their original semimajor axes, while almost $85 \%$ were tidally trapped. However, practically all had pericentric distances smaller than the Roche radius and were disrupted, so the overall formation efficiency of HJs was actually very low, of the order of $\simeq 0.7\%$. 


A similar result was found for orbits initially retrograde with respect to the perturber. In this case the tidal trapping appears to have almost the same efficiency; only $\simeq 7 \%$ remained near their original semimajor axis) while almost the same percentage as in the prograde case of the initial conditions lead to the formation of stable HJs. Finally, initial conditions retrograde with respect to the perturber but outside the Lidov-Kozai libration domain suffered a similar fate as those in direct orbits, with most of them avoiding tidal trapping. However, the effectivity of the formation of stable HJs is damped to $0.2 \%$ as expected, because of the planet being outside the Lidov-Kozai resonance. So, it appears that initial conditions with $I_{\rm mut} > 90^\circ$ seemed to be slightly more effective in forming stable HJs than those in direct orbits. 

Although the Lidov-Kozai mechanism proved to be inefficient in generating stable HJs, the final distributions of these bodies in the $(e_p,a_p)$ plane are very similar to that presented in Beaug\'e \& Nesvorn\'y (2012). Practically all form a kind of ``cascade'' or correlated distribution with $e_p \sim 0.8$ at $a_p = 0.2$ AU and ending with circular orbits at $a_p \simeq 0.03$ AU. The location of this cascade depends on the tidal parameter $Q'_p$ as well as the integration time, shifting towards larger semimajor axis for longer timescales or smaller values of $Q'_p$. 

The final distribution in the $(I_p,a_p)$ plane, however, presents significant differences with that obtained from planet-planet scattering. There is a noticeable lack of planets in low inclination orbits (both direct and retrograde), and the distribution appears almost symmetric with respect to polar orbits. Although Naoz et al. (2012) found a more uniform distribution, both distributions are in contrast with the observed distribution obtained from data of real Hot Jupiters. Although there are a few cases in which the initial conditions flip from direct to retrograde orbits (or vice-versa), most of the final distribution retains the original direction of the orbital angular momentum.

\begin{figure}[!th]
\centerline{\includegraphics*[width=12.0cm]{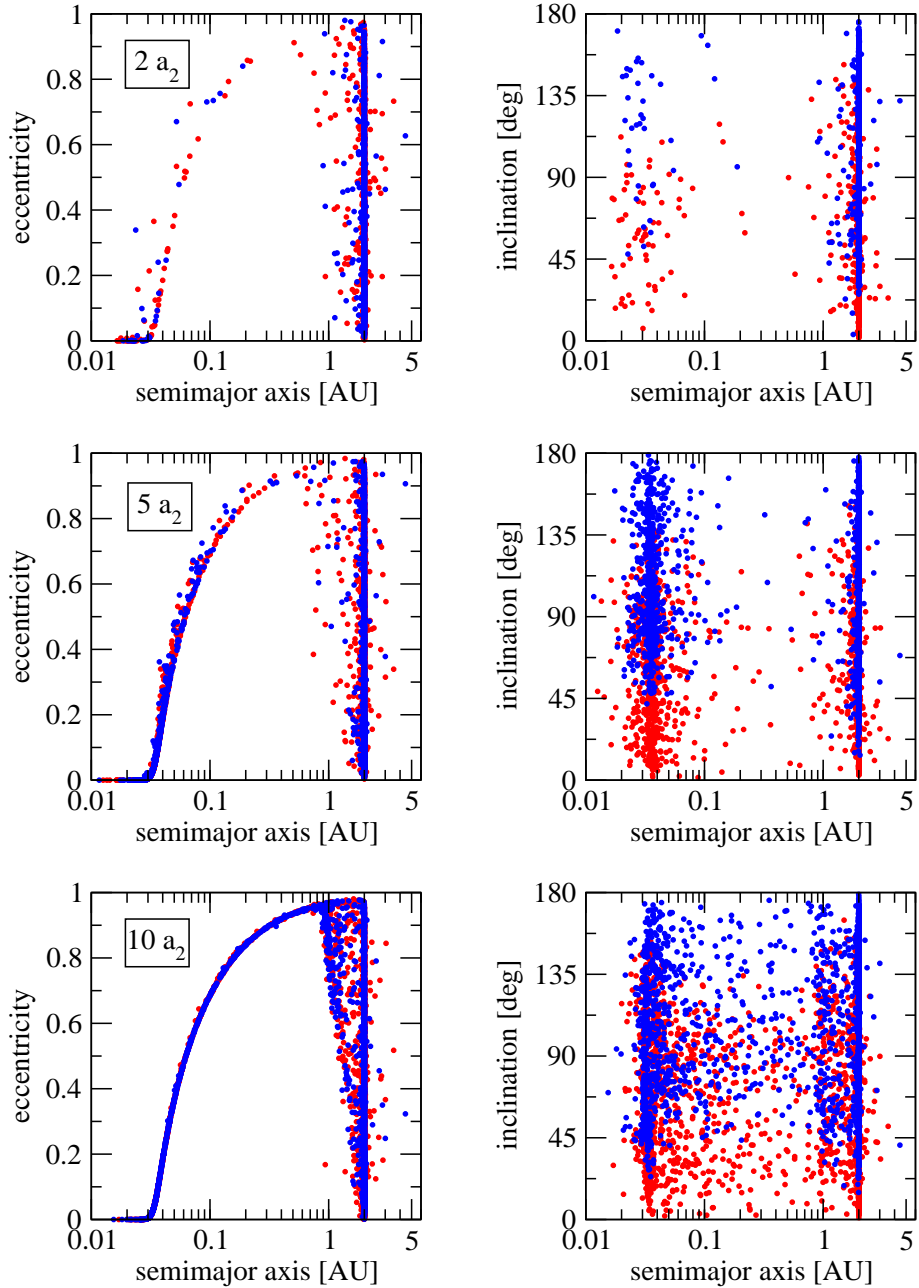}}
\caption{Final distribution of initial conditions with $I_{\rm mut} \in [40^\circ,140^\circ]$ where the semimajor axis of the binary was increased by a factor $N$ (i.e. $a_B \rightarrow N a_B$). Top frame shows results for $N=2$, middle for $N=5$ and bottom plot for $N=10$. Total integration time was $1$ Gyr. Color code is the same as previous figure.}
\label{fig3}
\end{figure}

As shown in Figure \ref{fig1}, most of our stellar companions have semimajor axis $a_B \sim 10-30$ AU, generating a very compact binary system (Although observations of planets in tight binary systems seem to show a minimum separation of 20 AU for planetary formation, we carried out our simulations for separations smaller than 20 AU as a matter of completeness). To analyze how the final distributions change with the stellar separation, we repeated the simulations applying a multiplicative factor $N$ to the initial value of $a_B$ in our initial conditions. Results are shown in Figure \ref{fig3} for $N=2$, $5$ and $10$, while the statistics of the outcomes are summarized in Table \ref{tab2}. 
In these runs we have only considered initial conditions within the Lidov-Kozai resonance domain
($I_{\rm mut} \in [40^\circ,140^\circ]$).

\begin{table}
\begin{center}
\begin{tabular}{rccc}
\hline
  $N$   &   Lost     &     HJs     &    Others   \\ 
\hline
  $2$   & $55.5 \%$  &   $1.5 \%$  &  $43.0 \%$  \\ 
  $5$   & $16.7 \%$  &  $20.7 \%$  &  $62.6 \%$  \\ 
  $10$  &  $0.8 \%$  &  $28.7 \%$  &  $70.5 \%$  \\ 
\hline
\end{tabular}
\end{center}
\caption{Outcomes of the simulations, as function of the separation factor $N$ of the binary companion.}
\label{tab2}
\end{table}

On one hand, we note a significant growth in the proportion of stable HJs as function of $N$, which increase from $1.5 \%$ for $N=2$ to almost $30 \%$ for wide binaries ($N=10$). Much of this new population is fed from planets that were tidally disrupted for lower values of $N$. Thus, wider binaries lower the number of planets engulfed by the star, many of whom are able to become tidally trapped at larger orbital distances and thus remain dynamically stable. This result is in agreement with those of Naoz et al. (2012), although our simulations cover a much greater range of semimajor axes.

The distribution in the $(e_p,a_p)$ and $(I_p,a_p)$ planes, however, shows little change as function of $N$. HJs still show a cascade-type distribution in eccentricity as function of the semimajor axis, although there appears to be a stronger correlation between both elements for wider binaries. The distribution of final inclinations still seems to favor values near polar orbits, and most of the initial orbital orientations are preserved after the planets become HJs.

\begin{figure}[t!]
\centerline{\includegraphics*[width=16.0cm]{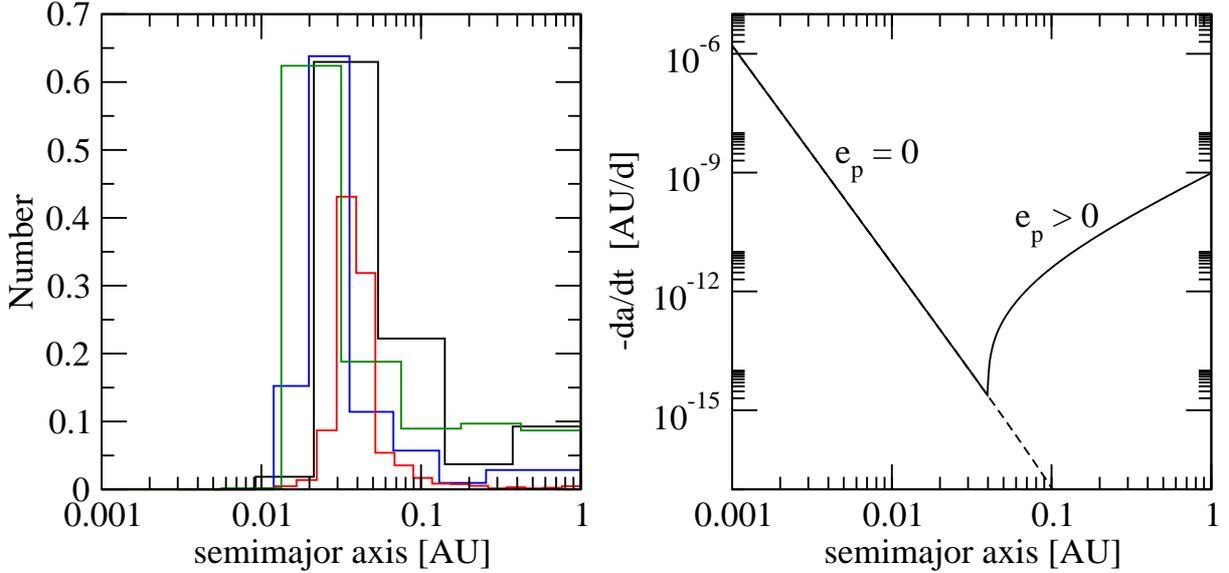}}
\caption{Left: Final distribution of simulated HJs as function of semimajor axis, for different values of $N$ (Black: $N=1$, Blue: $N=2$, Red: $N=5$, Green: $N=10$). The peak near $0.03$ AU corresponds to the 3-day pile-up in solar-type stars. Right: Value of the time derivative of the semimajor axis (sign changed), as function of $a_p$, for initial conditions at $1$ AU and $e_p=0.99$ (continuous line) and $e_p=0.1$ (dashed line).}
\label{fig4}
\end{figure}

In conclusion, although larger values of $a_B$ increase the efficiency of the formation process of hot planets, the final distribution of their orbits seems fairly invariant. A more detailed comparison is shown in the left-hand frame of Figure \ref{fig4}, where we plot the distribution of HJs in semimajor axis for different values of $N$. All show a peak near $a_p \simeq 0.03$ AU, which corresponds roughly to the 3-day pile-up in the case of solar-type central stars. Thus, the accumulation of observed giant planets around this value appears to be caused by the tidal trapping itself, and not by the driving mechanism (planet-planet scattering or Lidov-Kozai excitation). This fact can be observed in the right-hand frame of the same figure. Here we show the value of $-da_p/dt$, as a function of $a_p$, as calculated with our tidal model, initial values for $a_{p}$ and $e_{p}$ were chosen for ilustration purposes only. The continuous line corresponds to initial conditions equal to $a_p(t=0)=1$ AU and $e_p(t=0)=0.99$, while the dashed line corresponds to $e_p(t=0)=0.1$. As the semimajor axis decays with time, the evolution of the orbit occurs from right to left. The location of the cascade in the $(a_{p},e_{p})$ plane is actually determined by the tidal parameter. In this example we choose its value such that the limit for $e_p \rightarrow 0$ occurs close to the 3-day pile-up.

For initial quasi-parabolic orbits of the planet (corresponding to $e_{p}$ close to unity), during the first stage the planet's eccentricity remains finite and the planetary tide are much larger than the stellar counterpart. The magnitude of $|da_p/dt|$ decreases for smaller values of $a_p$, until it reaches a minimum for $e_p \sim 0$ (see Rodr\'{\i}guez \& Ferraz-Mello 2010). From this point onwards, orbital decay continues, but now fueled by stellar tides, whose effect increases for smaller semimajor axis. The value of the semimajor axis for which  $e_p \simeq 0$ is a (very pronounced) minimum of $|da_p/dt|$, and implies that bodies will remain near this spot for very long timescales. 

However, this minimum is only noted when tidal trapping occurs at high initial eccentricities. As noted in the dashed line, for low initial eccentricities ($e_p(t=0)=0.1$), $e_p$ is very quickly damped and the planet reaches the hot-planet region already in almost circular orbit and suffers a monotonic increase in the magnitude of orbital decay for all values of $a_p$.

Thus, the so-called 3-day pile-up appears to be both a consequence of tidal evolution and initial conditions with very high eccentricities, and constitutes observable evidence of tidal trapping. Moreover, the value of $a_p$ associated to this accumulation may be used to set the tidal parameter $Q'_p$, as shown by Beaug\'e \& Nesvorn\'y (2012). 

It should be pointed out that through-out the whole set of simulations the values of the masses of the bodies were held fixed at the masses of the bodies which form the actual $\gamma$-Cephei system. It is important to state that our results are valid only for the mass ratios used in this work. Although, there is no reason to believe that this particular selection of mass ratios are preferred in order to reproduce the same orbital characteristics that were obtained from planet-planet scattering and also some of the properties of the observed population of HJs. We suppose that these values of the masses of the bodies are sufficiently representative for the study of the Lidov-Kozai trapping as a mechanism for the HJ formation.

\section{The Cascade}

As with the case of planet-planet scattering, we found that the final distribution of HJs in the $(e_{p},a_{p})$ plane shows a strong correlation between those parameters and roughly resembles a cascade. Although its shape is similar to a curve of constant orbital angular momentum, both are not equivalent. This raises the questions as to what mechanism causes the appearance of the cascade and on which parameters it depends.

\begin{figure}[t!]
\centerline{\includegraphics*[width=16.0cm]{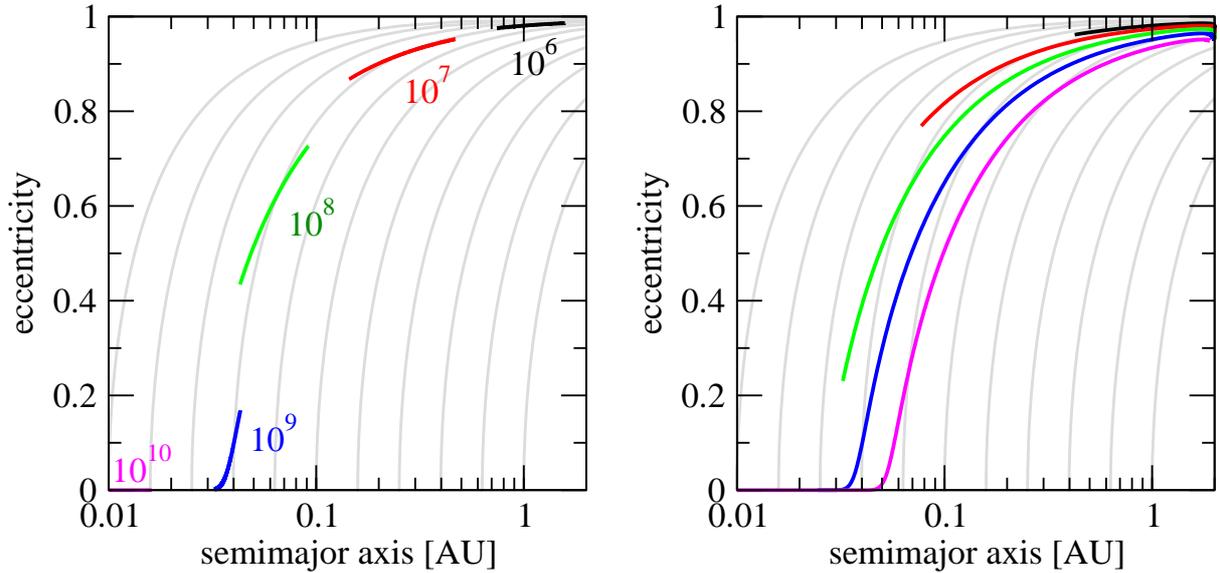}}
\caption{Simulations of tidal evolution of ensembles of initial conditions in the 2-body problem. Colors show these sets for different times (values in years). The gray curves correspond to  constant orbital angular momentum. Left: Initial semimajor axes $a_p \in [1.8,2.4]$ AU, and all eccentricities $e_p=0.99$. Right: All initial semimajor axis chosen equal to $a_p=2$ AU and eccentricities in the interval $e_p \in [0.952,0.992]$.}
\label{fig5}
\end{figure}

We will concentrate on the dynamical evolution of the planet after it has been tidally trapped, and can therefore disregard the gravitational perturbations from the binary. This simplifies the discussion, since we can study only the tidal effects in the 2-body problem. However, it is important to stress that this approximation is only employed here; all our previous simulations included the gravitational effects of the binary. Let us define a set of initial conditions in the $(e_p,a_p)$ plane defined by the intervals $a_p \in [a_0-\Delta a_p,a_0+\Delta a_p]$ and $e_p \in [e_0-\Delta e_p,e_0+\Delta e_p]$. 

If we choose $\Delta a_p = \Delta e_p = 0$, all initial conditions will define a single point and evolve simultaneously along a curve of constant angular momentum $G \propto \sqrt{a_p(1-e_p^2)}$. If we assume that the initial time is different for each initial condition, then the initial configuration will disperse, but still along a curve corresponding to a single value of $G$. This, however, does not correspond to the ``cascade'' type shape discussed above, which encompasses a finite interval of values of $G$.

Let us now consider $\Delta e_p = 0$ but $\Delta a_p = 0.2$ AU, generating an ensemble of initial conditions with a finite spread in semimajor axis but with the same eccentricity. this will show us how planets with different semimajor axis will cover the cascade. The evolution of such a set is depicted in the left-hand frame of Figure \ref{fig5}, where each color corresponds to a different time (its value given in years). For comparison, curves of constant $G$ are shown in gray. Even though we chose a fairly significant spread $\Delta a_p$, the ensemble still evolves as a close pack with little relative dispersion. Although its shape resembles a segment of the cascade, it never completes the figure, and again does not seem to explain the observed distribution.

The right-hand frame shows the results of a new set of initial conditions, this time with $\Delta a_p=0$ and $e_p \in [0.952,0.992]$, which will now ilustrate how  planets with different eccentricities cover the cascade. Again the color code shows the shape of the ensemble for different times, and the individual values are the same as those in the other graph. Even though we adopted a small value for $\Delta e$, since we are in the realm of almost parabolic orbits it generates a large dispersion in pericentric distance. This has two consequences: first, a large spread in $G$ and, second, very different tidal decay timescales. So, while those initial conditions with the lower eccentricities suffer almost no change in the semimajor axis throughout the simulation (even for times $\sim 10^{10}$ years), those with higher values of $e_p$ suffer a much quicker orbital decay. The result is a cascade covering a significant interval in values of $G$ and $e_p$, and very similar to that obtained from both our planet-planet scattering and Lidov-Kozai trapping experiments. 

Summarizing, the cascade-like structure in the $(e_p,a_p)$ plane obtained for HJs in the tidal trapping simulations seems primarily caused by small differences in the planetary eccentricity at the moment of tidal capture. Differences in semimajor axis cause little effect, while differences in time of capture will also be slight. For example, there is not much change in the location of the cascade from $t=10^9$ to $t=10^{10}$ yrs, and any difference in the initial time will be constrained by these two curves.

\section{Conclusions}

In this paper we have analyzed the formation scenario of HJs based on tidal trapping from quasi-parabolic orbits excited by Lidov-Kozai resonances with a binary stellar component. We have used the same tidal and gravitational model as in Beaug\'e \& Nesvorn\'y (2012), where we discussed the same problem but assuming planet-planet-scattering as the catalyst. Our aim has been two-fold. First, compare the final distribution of HJs predicted by both mechanisms using similar dynamical tools, models and parameters. Second, try to understand the origin of some of the observed characteristics of the simulations and the observed planets.

We have found that several of the final orbital properties of the simulated HJs (as well as the real bodies) are mainly caused by the process of tidal trapping, and independent of the excitation mechanism. This includes both the distribution in the $(e_p,a_p)$ and the 3-day pile-up. Both may be considered as observational evidence that a significant portion (or even most) of the observed HJs originated from tidal trapping and not from smooth disk-induced migration. 

\begin{figure}[t!]
\centerline{\includegraphics*[width=16.0cm]{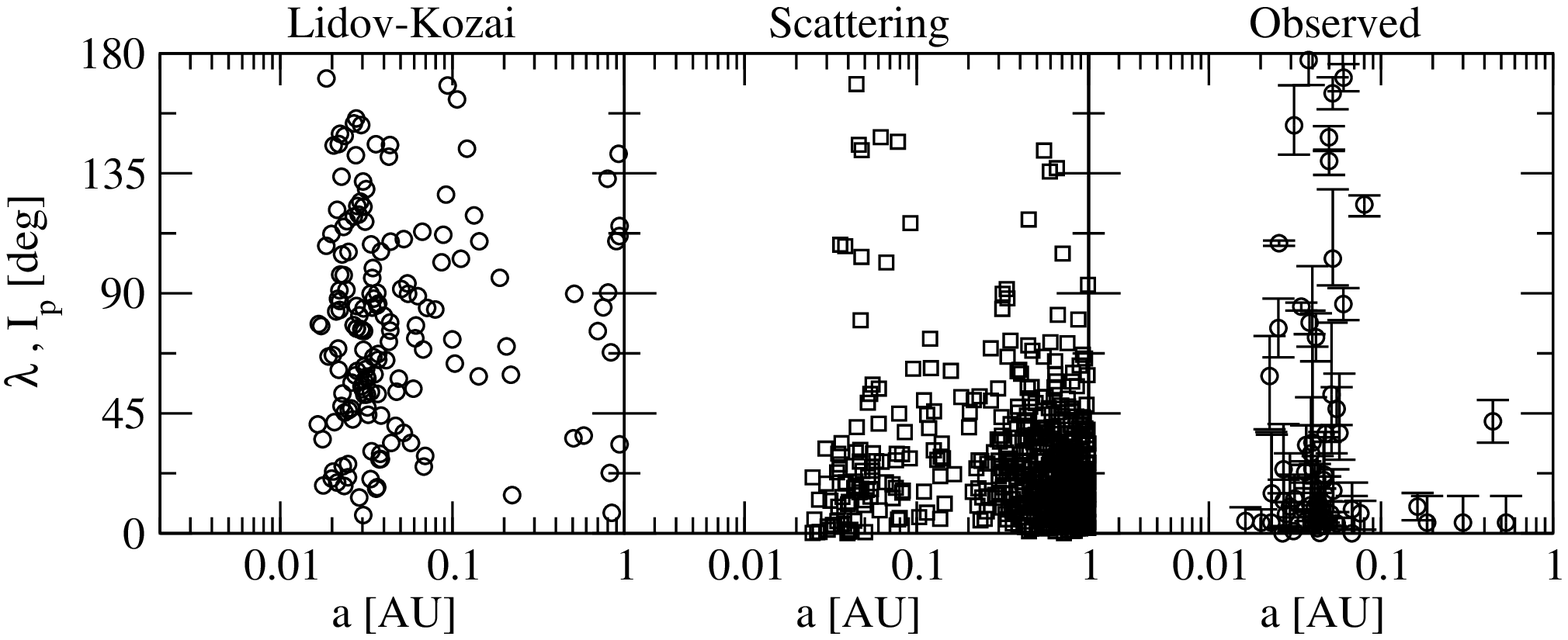}}
\caption{Distribution of the final inclination between the planet and the stellar equator, obtained from our Lidov-Kozai simulations (left plot), as compared with the results from planet-planet scattering experiments (Beaug\'e \& Nesvorn\'y 2012) (center plot) and with the distribution of the sky-projected misalignment angle $\lambda$ for observed HJs (right plot).}
\label{fig6}
\end{figure}

However, we have also found that the final distribution of the inclinations is different with respect to those obtained from planet-planet scattering experiments. Figure \ref{fig6} shows a comparison of both, together with the observed distribution of the sky-projected misalignment angles currently available for HJs. While the Lidov-Kozai mechanism shows a notorious absence of low-inclination orbits, the distribution obtained from scattering and the observed values show a greater proportion of orbits nearly coplanar with the stellar equator. 

The reason behind this difference may lie in the excitation mechanism itself. In planet-planet scattering there is no direct correlation between the excitation in eccentricity and that in inclination, and it is possible (albeit likely) that high-eccentricity orbits remain with low inclinations (e.g. below $\sim 45^\circ$). Thus, in this scenario, most of the HJs will preferably remain aligned with the equator of the central star, as observed in real planets. In Lidov-Kozai trapping, formation of HJs require a high inclination with respect with the perturbing mass. Unless the binary is conveniently located in a polar orbit with respect to the equator of $m_A$, the final inclinations of the hot planets will show a more random distribution, with no preference for almost aligned orbits. 

Naoz et al. (2012) proposed to solve this issue arguing in favor of the existence of a second population of HJs generated by smooth disk-induced planetary migration and, consequently, containing low inclinations. However, as noted from Figure \ref{fig4}, this second population should not show a 3-day pile-up in orbital periods, and thus cannot explain the fact that most of the real HJs in this pile-up show small misalignment angles. Planet-planet scattering suffers from none of these limitations, and thus appears to be more consistent with the distribution of real planets. Even so, it is indeed possible that all three proposed mechanisms (Lidov-Kozai, scattering and smooth planetary migration) could have contributed to the complete sample of known HJs. Only future work, including combined scenarios and additional data, will allow us to speculate as to the effective role of each of them.

\vspace*{0.5cm}
\section*{Acknowledgements}
This work has been partially funded by CONICET and SECYT/UNC. The authors would like to express their gratitude to N. Haghighipour for the invitation us to participate in this special issue, and to both referees for valuable suggestions.

\vspace*{0.5cm}
\section*{References}
\vspace*{0.5cm}

\noindent
Beaug\'e, C., Nesvorn\'y, D. 2012. ApJ 751, 119.

\noindent
Ben\'{\i}tez-Llambay, P., Masset. F., Beaug\'e, C. 2011. A\&A 528, A2.

\noindent
Correia, A.C.M., Laskar, J., Farago, F., Bou\'e, G. 2011. CeMDA 111, 105.

\noindent
Haghighipour, N. 2010, Planets in Binary Star Systems, Springer, New York

\noindent
Harrington, R.S. 1968. AJ 73, 190.

\noindent
Innanen, K. A., Zheng, J. Q., Mikkola, S., Valtonen, M. J. 1997. AJ 113, 1915.

\noindent
Juric, M., Tremaine, S. 2008. ApJ 686, 603.

\noindent
Kozai, Y. 1962. AJ 67, 591.

\noindent
Laskar, J., Bou\'e, G. 2010. A\&A 552, A60.

\noindent
Libert, A.-S., Henrard, J. 2007. Icarus 191, 469. 

\noindent
Lin, D., Bodenheimer, P. Richardson, D. 1996. Nature, 380, 606.

\noindent
Lidov, M. L. 1961. Isk. Sput. Zemli, 8, 119.

\noindent
Mart\'i, J., Beaug\'e, C. 2012. A\&A 544, A97.

\noindent
Nagasawa, M., Ida, S. 2011, ApJ, 742, 72.

\noindent
Nagasawa, M., Ida, S., Bessho, T. 2008, ApJ, 678, 498.

\noindent
Naoz, S., Farr, W.M., Lithwick, Y., Rasio, F.A., Teyssandier, J. 2011. Nature, 473, 187.

\noindent
Naoz, S., Farr, W.M., Lithwick, Y., Rasio, F.A. 2012. ApJL, 754, L36.

\noindent
Rasio, F., Ford, E. 1996.  Science, 74, 954-956.

\noindent
Rodr\'{\i}guez, A., Ferraz-Mello, S. 2010.  EAS Publications Series, 42, pp 411.

\noindent
Winn, J.N., Fabrycky, D., Albrecht, S., Johnson, J.A. 2010. ApJL 718, L145.

\end{document}